\documentclass[11pt]{article}
\usepackage[margin=1in,pdftex]{geometry}

\usepackage{graphicx}
\usepackage[bottom]{footmisc}
\usepackage{listings}
\usepackage{enumitem}

\usepackage{wrapfig}
\usepackage{ragged2e}

\usepackage{amsmath}
\usepackage{multicol}
\usepackage{lipsum}
\usepackage{hyphenat}
\PassOptionsToPackage{hyphens}{url}
\usepackage{hyperref}
\usepackage{url}

\usepackage{rotating}

\newcommand*{\email}[1]{\small{\texttt{#1}}}

\renewcommand{\footnoterule}{%
  \kern -3pt
  \hrule width \textwidth height 0.5pt
  \kern 2pt
}

\title{\bfseries
An Interactive, Graphical\\CPU Scheduling Simulator\\for Teaching Operating Systems}

\author{Joshua W. Buck and Saverio Perugini\\
Department of Computer Science\\
University of Dayton\\
300 College Park\\
Dayton, Ohio\ \ 45469--2160\ \ USA\\
Tel: +001 (937) 229--4079\\
E-mail: \email{joshua.buck993@gmail.com}~~~~~\email{saverio@udayton.edu}\\
Demo Site: \url{https://cpudemo.azurewebsites.net}\\
Project Site: \url{http://sites.udayton.edu/operatingsystems}}

\begin{document}
\sloppy
\maketitle

\thispagestyle{empty}

\begin{abstract} We present a graphical simulation tool for visually and
interactively exploring the processing of various events handled by an
operating system when running a program.  Our graphical simulator is available
for use on the web and locally by both instructors and students for
purposes of pedagogy.  Instructors can use it for live demonstrations of course
concepts in class, while students can use it outside of class to explore the
concepts.  The graphical simulation tool is implemented using the React library
for the fancy \textsc{ui} elements of the \texttt{Node.js} framework and is
available as a single page web application at \url{https://cpudemo.azurewebsites.net}.
Assigning the development of the underling text-based simulation engine, on
which the graphical simulator runs, to students as a course project is also an
effective approach to teach students the concepts.  The goals of this paper are
to showcase the demonstrative capabilities of the tool for instruction, share
student experiences in developing the engine underlying the simulation, and to
inspire its use by other educators.  \end{abstract}

\paragraph{Keywords:}
\texttt{Node.js};
operating systems pedagogy;
process scheduling;
React library;
semaphore processing.

\section{Introduction}

We present a graphical simulation tool for visually and interactively exploring
the processing of a variety of events handled by an operating system when
running a program~\cite{Buck:2019:IGS:3287324.3293756}.  Our tool graphically
demonstrates a host of concepts of preemptive, multi-tasking operating
systems, including scheduling algorithms, \textsc{i/o}
processing, interrupts, context switches, task structures (i.e.,
process control blocks), and semaphore
processing~\cite{osc10}.

\begin{figure}[t]
\centering
\resizebox{\textwidth}{!}{
\includegraphics[scale=1.0]{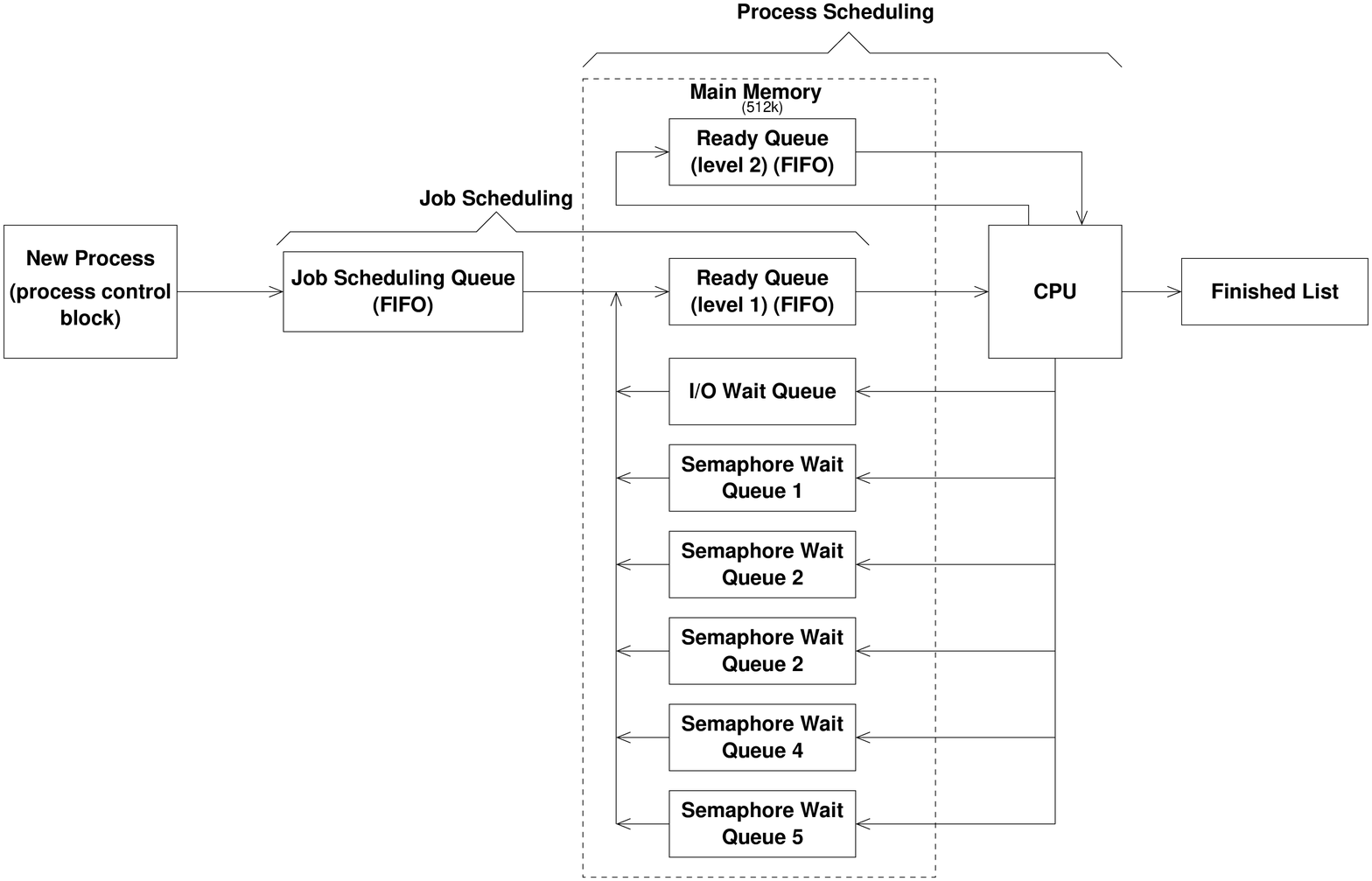}}
\caption{Architectural view of the underlying simulator depicting the transitions
processes can take as they move throughout the various queues and the 
\textsc{cpu} of the system.}
\label{fig:arch}
\end{figure}

Our graphical tool was designed to run on top of a solution to a text-based
course programming project---here after referred to as the \textit{underlying
simulation engine}---in which students design and implement a system that
simulates some of the job and \textsc{cpu} scheduling, and semaphore processing
of a time-shared operating system (\textsc{os}).  The complete project
specification for the underlying simulation engine is available at
\url{http://perugini.cps.udayton.edu/teaching/courses/Fall2015/cps356/projects/p2.html}.\footnote{A
specification for the underlying simulation engine without system semaphores is
available at
\url{http://perugini.cps.udayton.edu/teaching/courses/Fall2019/cps356/\#midterm}.}
The architectural view of the underling simulation engine, which also serves to
convey some of the project/system requirements, is shown in
Figure~\ref{fig:arch}. While processes from the second-level ready queue
receive a larger quantum than those from the first-level ready queue (300
versus 100), processes on the first-level ready queue always have priority over
processes on the second-level ready queue.  In Figure~~\ref{fig:arch}, notice
that once a process from the first-level ready queue has received a full
quantum it is promoted to the second-level ready queue.  However, if a process
from the second-level ready queue requires \textsc{i/o}, it is demoted back to
the first-level ready queue once it receives that \textsc{i/o}.   Students are
familiar with the concepts of job and processing scheduling, non-preemptive and
preemptive scheduling algorithms, as well as semaphores prior to working on
this project.  See
\url{http://perugini.cps.udayton.edu/teaching/courses/cps346/lecture_notes/scheduling.html}
(scheduling) and
\url{http://perugini.cps.udayton.edu/teaching/courses/cps346/lecture_notes/semaphores.html}
(semaphores) for more information.

While demonstrating \textsc{os} concepts using physical computer hardware and
real operating systems is effective, a software simulation is less expensive to
develop and more easily configurable.  For instance, users of our tool have
control over both the time-based events and the parameters of the system (e.g.,
quantum size and main memory constraints) that are less
easily controllable at the
user level in a real computing system.  Moreover, a visual simulation
graphically reveals the internal processing of and event handling within an
\textsc{os} from which the user is typically shielded.
The ability to step through the handling of events (e.g., new process creation,
process termination, \textsc{i/o} completion, a context switch) enabled by our
tool in user-defined steps of \textsc{cpu} time is formative in students'
conceptualization, comprehension, and visualization of these complex processes
at work within an \textsc{os}.

The graphical simulation tool is implemented using the React library for the
fancy \textsc{ui} elements of the \texttt{Node.js} framework and is available
as a web application at \url{https://cpudemo.azurewebsites.net/}.
Our graphical simulator is available for use on the web as well as locally by
both instructors and students for purposes of pedagogy.  Instructors can use it
to for live demonstrations of course concepts in class, while students can use
it outside of class to explore the concepts.  Assigning the development of the
underling text-based simulation engine, on which the graphical simulator runs,
to students as a course project is also an effective approach to teach students
the concepts---more on this below.

\section{Simulation Details}

The top of the left-most column of the tool, shown in Figure~\ref{fig:demo1},
contains a list of incoming external events in the current session. Below the
incoming external event list is a list of jobs rejected by the system because
each requires more memory than the total system memory.  At the bottom of the
left-most column of the tool is the job scheduling queue
(in secondary memory), which lists all jobs
that are waiting for main memory to become available before they can be moved
to the ready queue. The middle columns of the tool contain the multi-level,
\textsc{fifo}  ready queue, the \textsc{i/o}  wait queue, and the semaphore
wait queues.
The right-most column of the tool contains a block representing the
\textsc{cpu}, and a list of the completed processes.  Above the \textsc{cpu},
the available system memory is shown.

The user can step through events in multiple ways. The top of the tool shows
the current simulation time, which the user can modify at any time with values between 0 and 30,000.
Alternatively, the user can use the slider bar handle to advance or subtract up
to 250 units of simulation time at once. Upon changing the time with the slider
bar, the handle resets to the middle position and the user can again advance or
subtract up to 250 units of time. There are also controls to allow the user to
step forward to the next or backward to the previous simulation event.
Finally, there is a button labeled `Complete Run' to fully run the simulator and
produce the output (i.e., a variety of turnaround and wait time statistics) of
the underlying text-based simulation engine in one stroke.

There are seven events in this simulator: four external events (i.e., given in
the incoming external event list) and three internal events\footnote{There are
really eight events because there is a display simulator status external event
which while relevant in the text-based interface to the underlying simulation
engine is not applicable to the graphical interface presented here.}.  The
four external events are a new job arrival, an \textsc{i/o} request, a
semaphore wait and a semaphore signal. The three internal events are process
termination, quantum expiration, and \textsc{i/o} completion.  If an external
and internal event collide (i.e., occur at the same time), the internal event
is processed first.
The events are automatically processed in the background and the event
navigation buttons allow a user to see every change that occurs in the
simulation. At any point, the user may click the button labeled `Reset Run'
to return the simulation time to 0. 

\subsection{Customization: Tuning the Simulation Parameters}

\begin{figure}[h!]
\centering
\resizebox{\textwidth}{!}{
\begin{tabular}{c}
\includegraphics[scale=0.20]{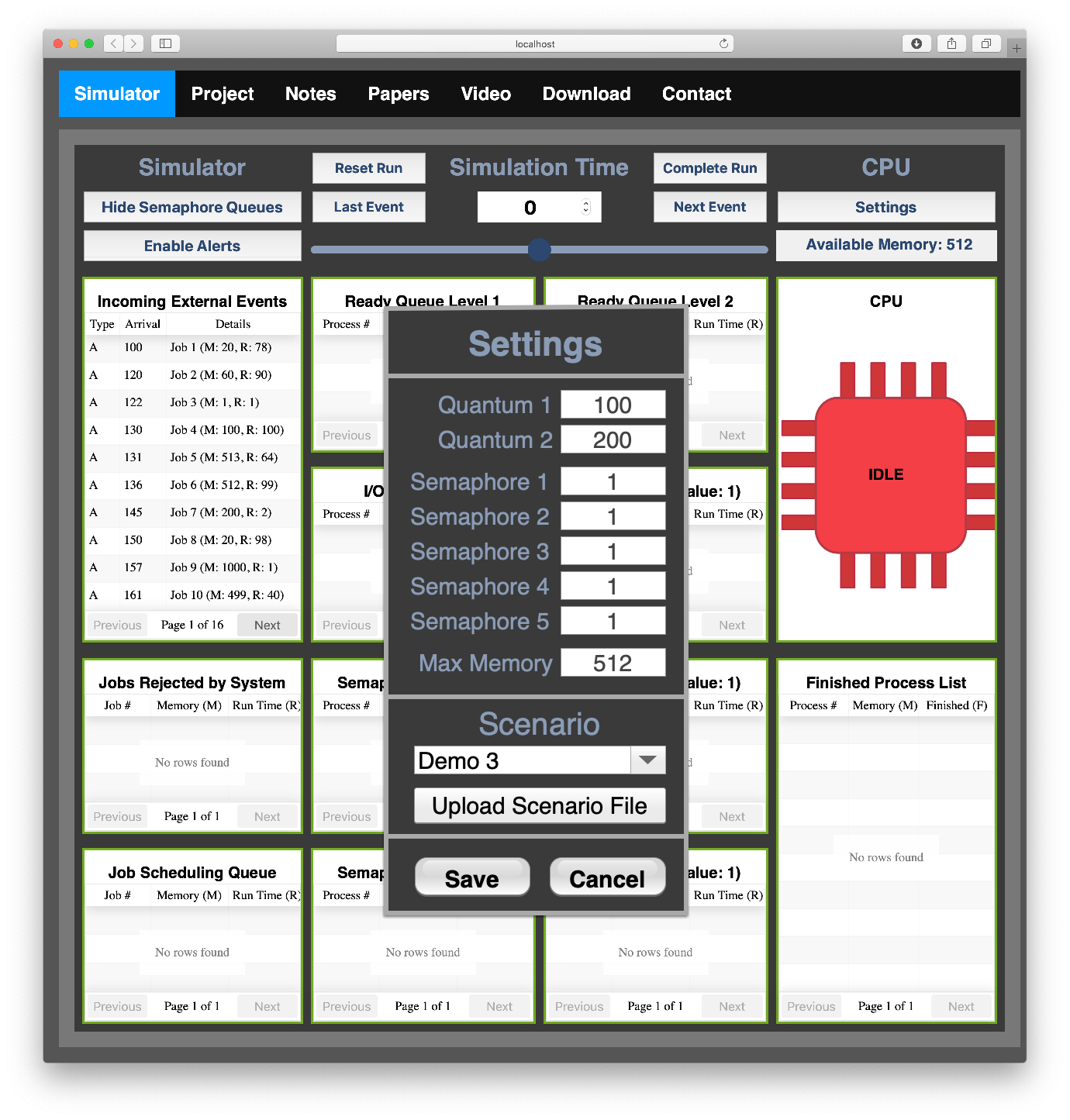}
\end{tabular}}
\caption{Sample screen, illustrating parameter tuning,
from the simulation tool (Screen 2 of 17).}
\label{fig:demo2}
\end{figure}

\begin{figure}[h!]
\centering
\resizebox{\textwidth}{!}{
\begin{tabular}{c}
\includegraphics[scale=0.20]{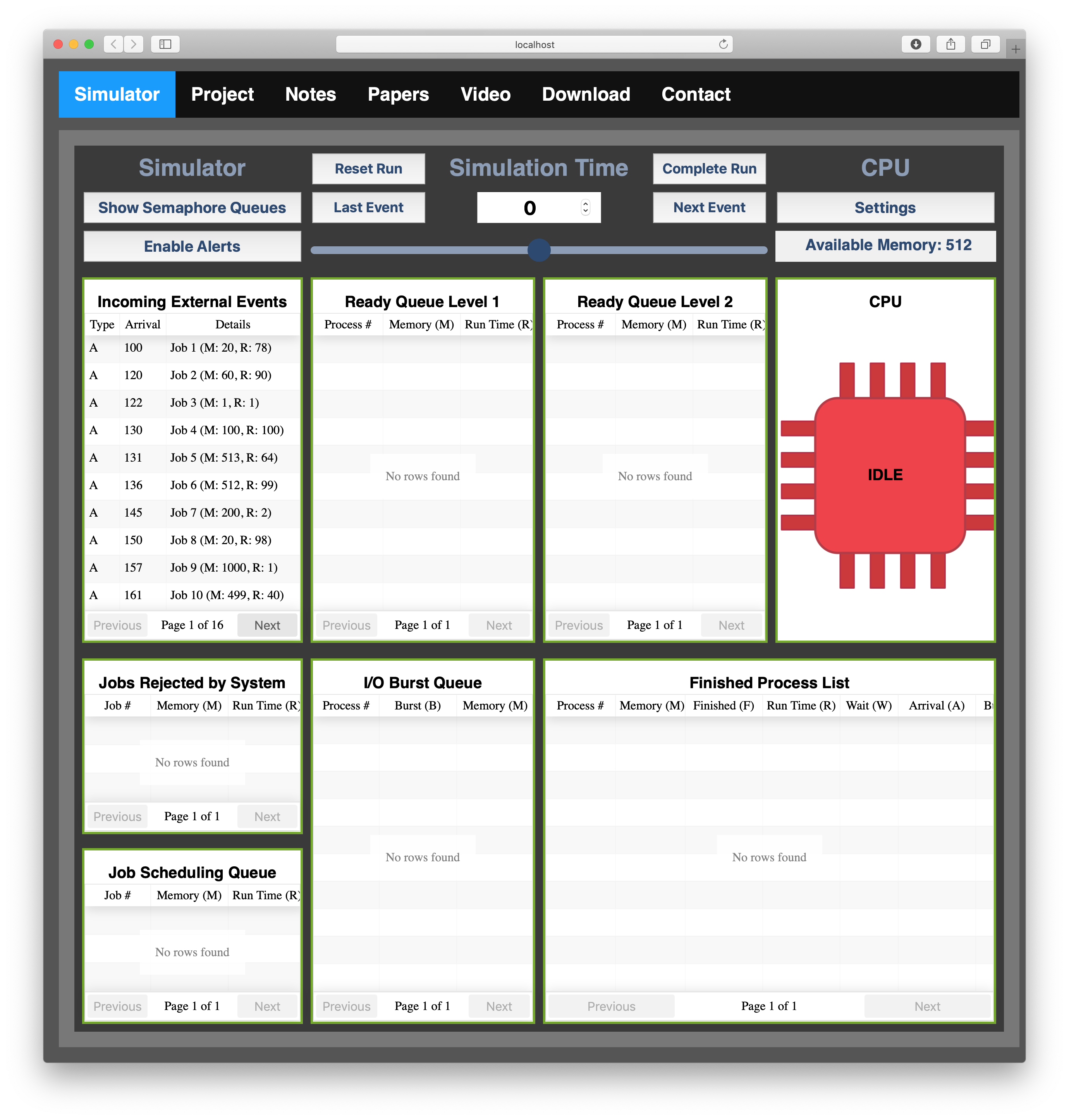}
\end{tabular}}
\caption{Sample simplified screen from the simulation tool (Screen 3 of 17).}
\label{fig:demo3}
\end{figure}

There are several ways in which a user can customize the simulation. The
button labeled `Settings' opens a dialog window, shown in Figure~\ref{fig:demo2}, with
several simulation variables that can be tuned.  The user can set the quantum
for each level of the \textsc{fifo} ready queue.  Setting the quantum of a
particular level of the ready queue to 0, sets the scheduling algorithm to be
`first-come, first-served' (\textsc{fcfs})---a non-preemptive algorithm---for
that level.  The user can also select a simulation scenario (i.e., a sequence
of incoming external events) from a drop-down menu of canned event scenarios.
Alternatively, a user can upload their own simulation scenario 
as text file.  Creating and
importing such custom sequences of simulation events is helpful for
demonstrating specific scenarios, especially event collisions.  The user can
also change the current values of any of the semaphores and set the maximum
memory available to the simulation. Jobs that require more memory than the
system maximum are rejected and placed in the list of jobs rejected by the
system. When a job is rejected by the system for any reason, an alert is
displayed in the tool. There is a toggle available for disabling or enabling
these alerts. Finally, another toggle is available that will simplify the tool
layout by hiding the semaphore queues and enlarging the other queues. The
simplified layout is shown in Figure~\ref{fig:demo3}.

\subsection{Example Simulation Scenario}

\begin{figure}[h!]
\centering
\resizebox{\textwidth}{!}{
\begin{tabular}{c}
\includegraphics[scale=0.20]{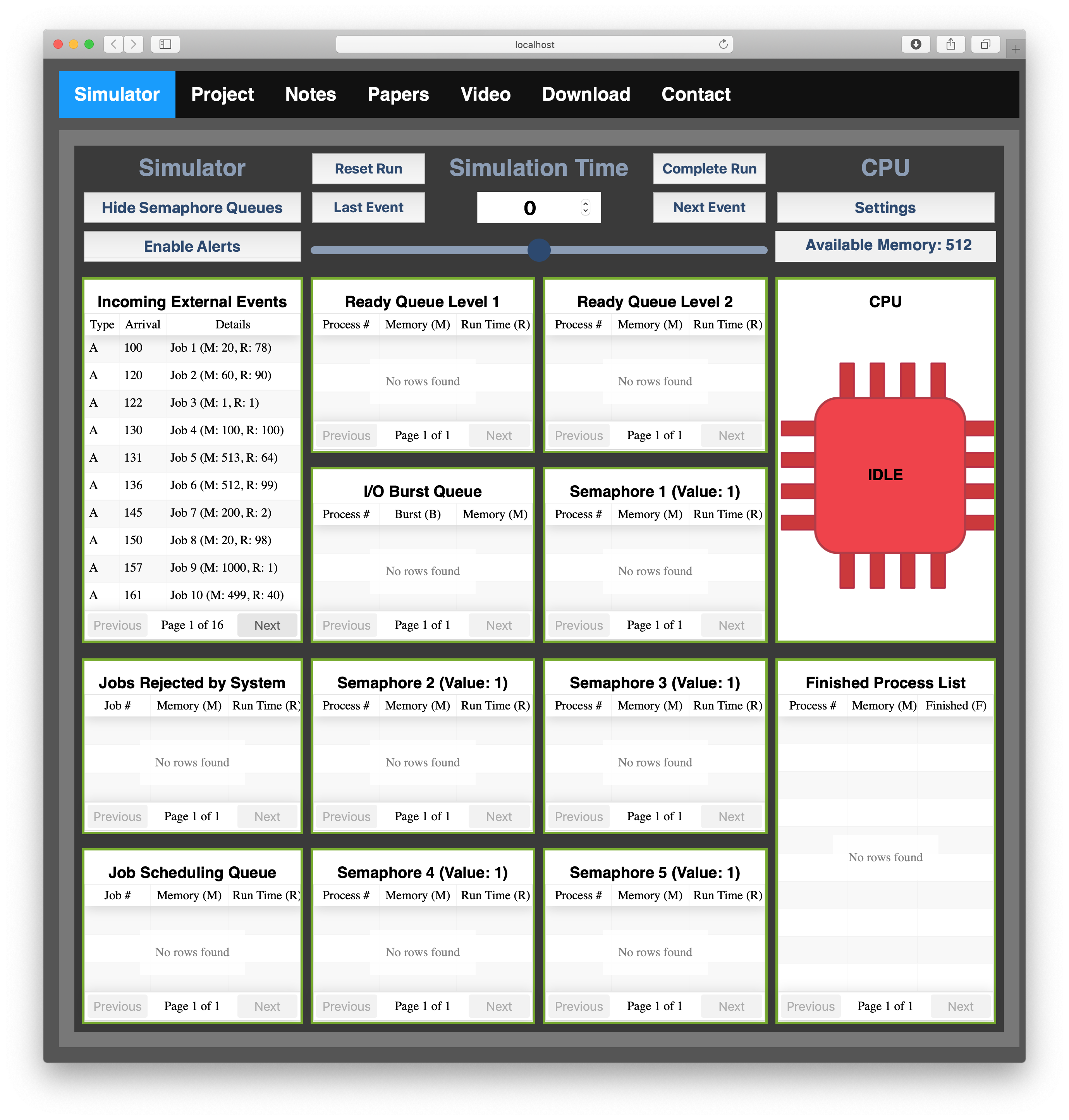}
\end{tabular}}
\caption{Sample screen from the simulation tool before any
event has occurred (Screen 1 of 17).}
\label{fig:demo1}
\end{figure}

\begin{figure}[h!]
\centering
\resizebox{\textwidth}{!}{
\begin{tabular}{c}
\includegraphics[scale=0.20]{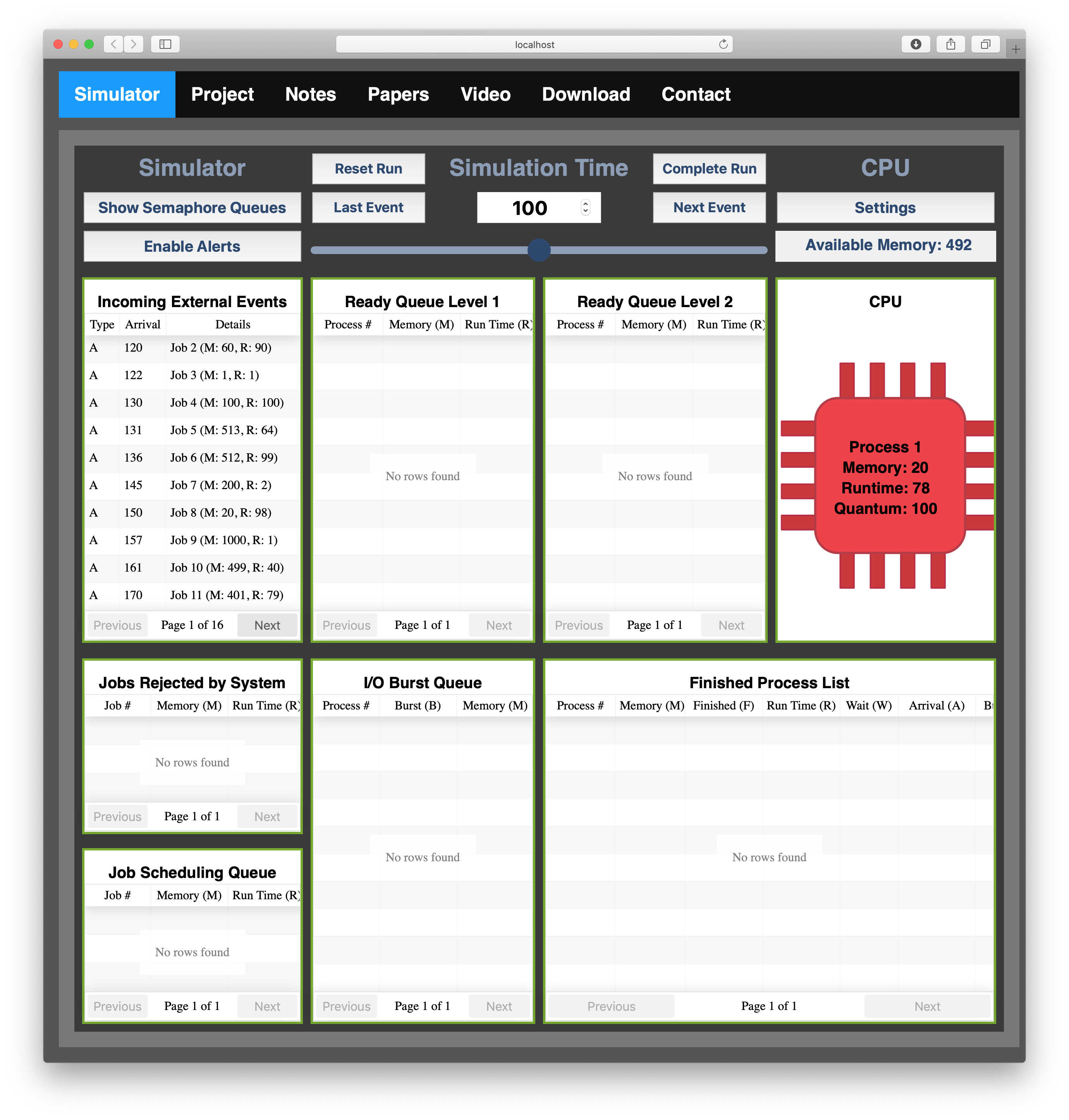}
\end{tabular}}
\caption{Sample screen from the simulation tool (Screen 4 of 17).}
\label{fig:demo4}
\end{figure}

\begin{figure}[h!]
\centering
\resizebox{\textwidth}{!}{
\begin{tabular}{c}
\includegraphics[scale=0.20]{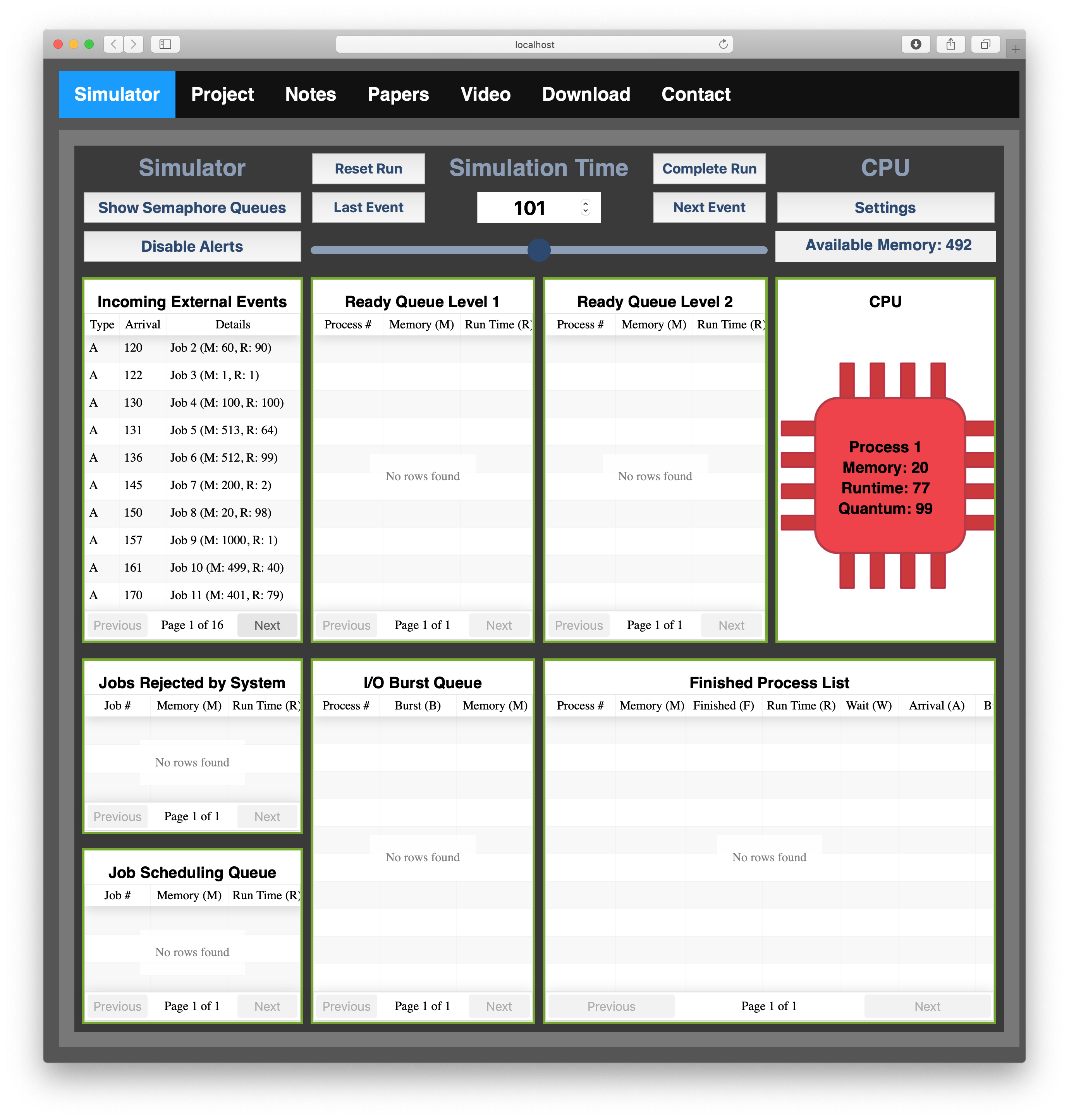}
\end{tabular}}
\caption{Sample screen from the simulation tool (Screen 5 of 17).}
\label{fig:demo5}
\end{figure}

Figure~\ref{fig:demo1} shows the system before any events occur. In
Figure~\ref{fig:demo4}, 100 units of time have elapsed and the first event
occurs---a new job arrival---identified by the letter `\textsc{a}'
in the first column
of the incoming external events list (see Figure~\ref{fig:demo1}).  Note that
the job id, and runtime and memory requirements are also provided in the
incoming external events list. The new job immediately moves into the job
scheduling queue, which triggers the job scheduling algorithm.  Since job 1
requires 20 units of memory and 512 units are available, job 1 is immediately
loaded into the first level of the ready queue. Since the \textsc{cpu} is idle
at this point, the arrival of a process in the ready queue triggers the
\textsc{cpu} scheduling algorithm, which moves process 1 from the ready queue
and onto the \textsc{cpu} with a quantum of 100 units of time. 
While this simulation does not currently factor in the time required for the
overhead of a \textsc{cpu} context switch, process 1 does not run until the
next clock cycle of the \textsc{cpu}. A summary of time stamp 100 is: the first
job arrived and
was instantaneously loaded onto the \textsc{cpu} (through the job
queue and first-level ready queue). At time 101, shown in
Figure~\ref{fig:demo5}, process 1 runs for 1 unit of the 78 required units of
time before completion and has a remaining quantum of 99 clock cycles.  Since
the remaining quantum for process 1 is 99 units of time, the process would
finish without time-slicing, but may require \textsc{i/o} or a
semaphore in the interim.  The incoming external events list in
Figure~\ref{fig:demo5} shows that
the next incoming external event---another new job
arrival---occurs at simulation time 120.

\begin{figure}[h!]
\centering
\resizebox{\textwidth}{!}{
\begin{tabular}{c}
\includegraphics[scale=0.20]{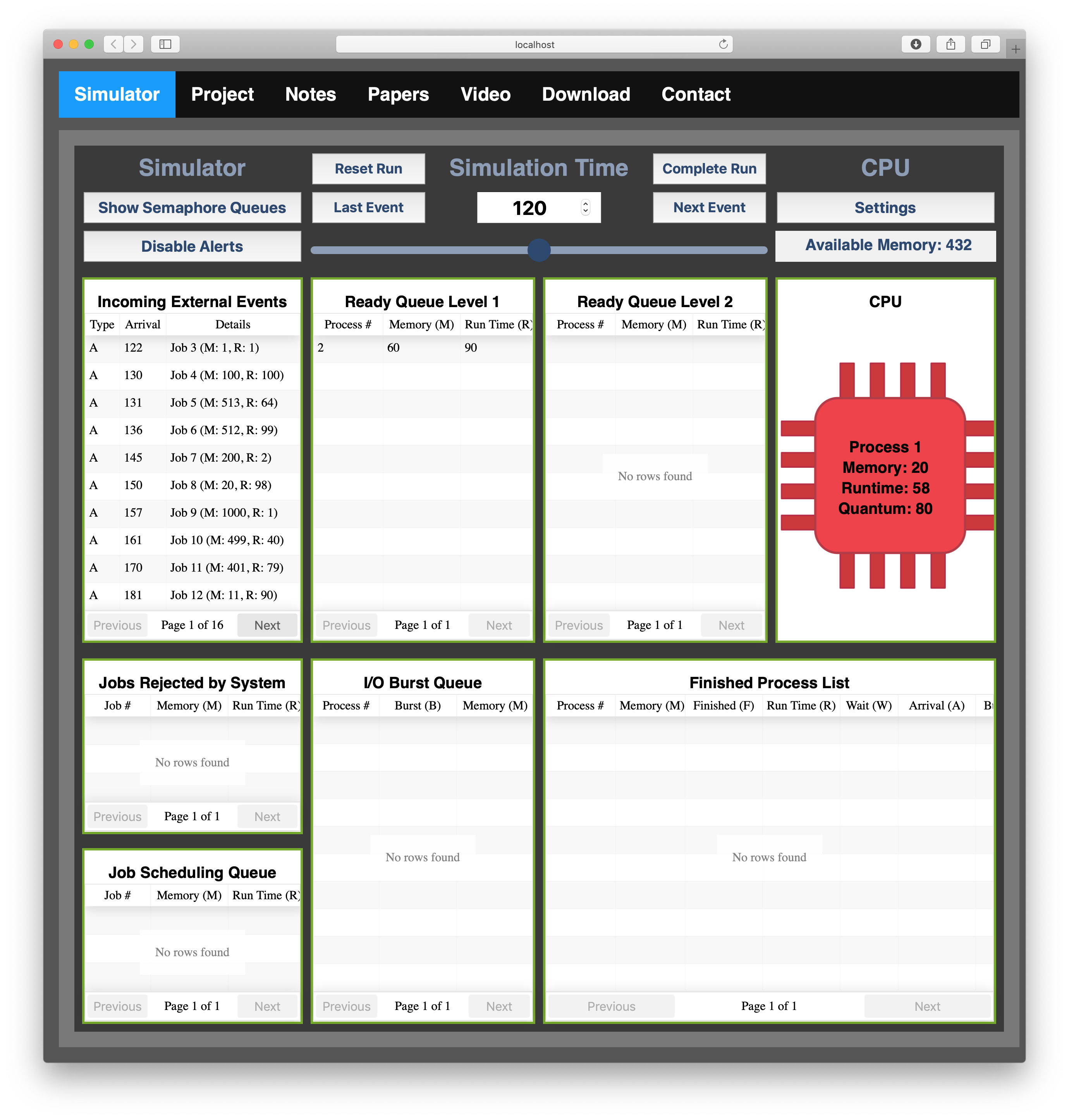}
\end{tabular}}
\caption{Sample screen from the simulation tool (Screen 6 of 17).}
\label{fig:demo6}
\end{figure}

Clicking the button labeled `Next Event' to the right of the simulation time
automatically advances the time to the next event. At this point, the next
event is the next incoming external event as opposed to an internal event. At
time 119, process 1 has run for a total of 19 units of time, and requires 59
additional units of time until completion.  During the next unit of time, shown
in Figure~\ref{fig:demo6}, a new job arrives.  It is moved into the job queue,
which triggers the job scheduling algorithm.  Since job 2 requires 60 units of
memory, and there are 492 units available, job 2 is immediately loaded into the
first level of the ready queue. Since the \textsc{cpu} is busy with process 1
at this point in time, process 2 must wait in the ready queue until the
\textsc{cpu} is available. 

\begin{figure}[h!]
\centering
\resizebox{\textwidth}{!}{
\begin{tabular}{c}
\includegraphics[scale=0.20]{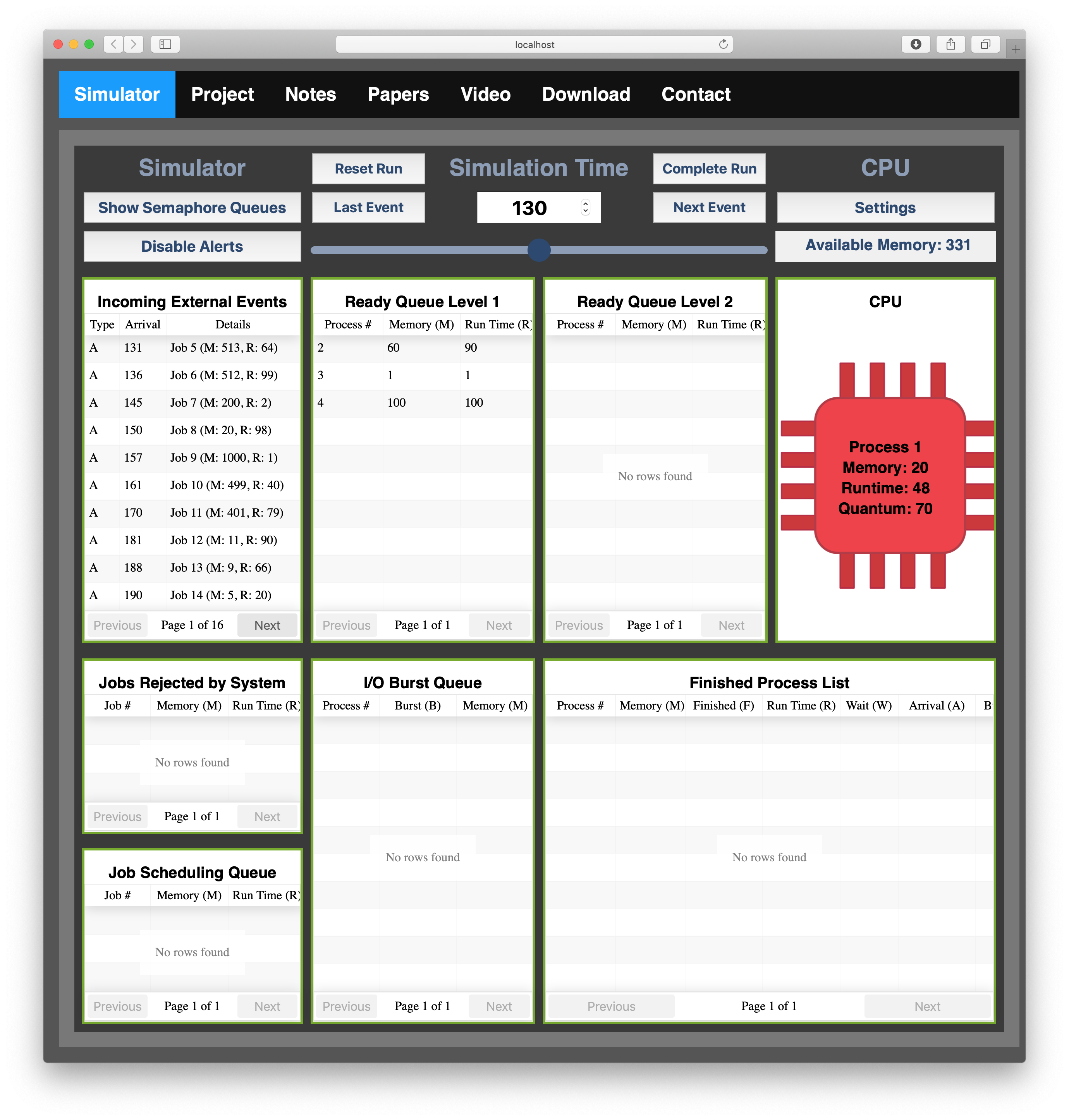}
\end{tabular}}
\caption{Sample screen from the simulation tool (Screen 7 of 17).}
\label{fig:demo7}
\end{figure}

\begin{figure}[h!]
\centering
\resizebox{\textwidth}{!}{
\begin{tabular}{c}
\includegraphics[scale=0.20]{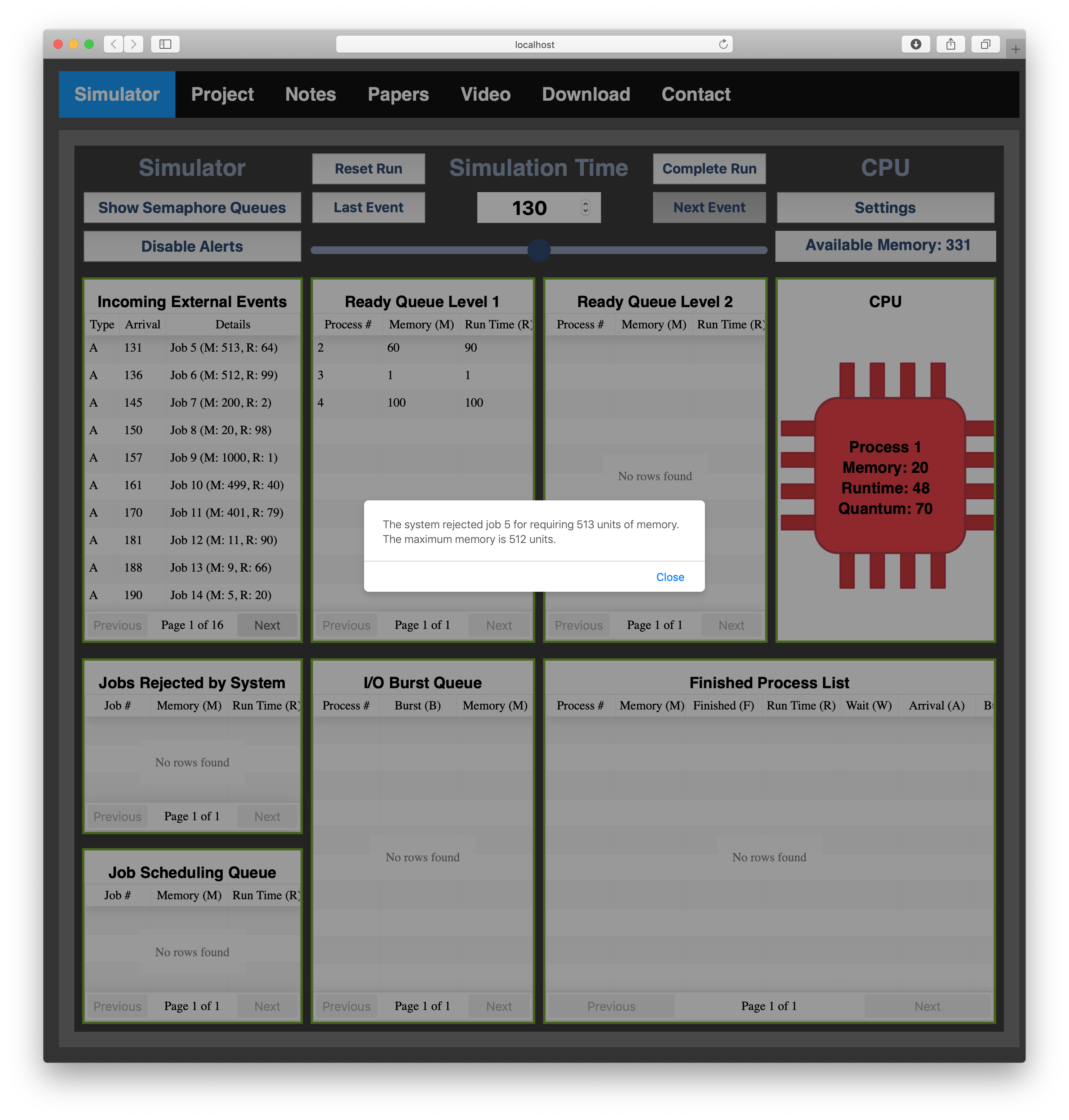}
\end{tabular}}
\caption{Sample screen from the simulation tool (Screen 8 of 17).}
\label{fig:demo8}
\end{figure}

\begin{figure}[h!]
\centering
\resizebox{\textwidth}{!}{
\begin{tabular}{c}
\includegraphics[scale=0.20]{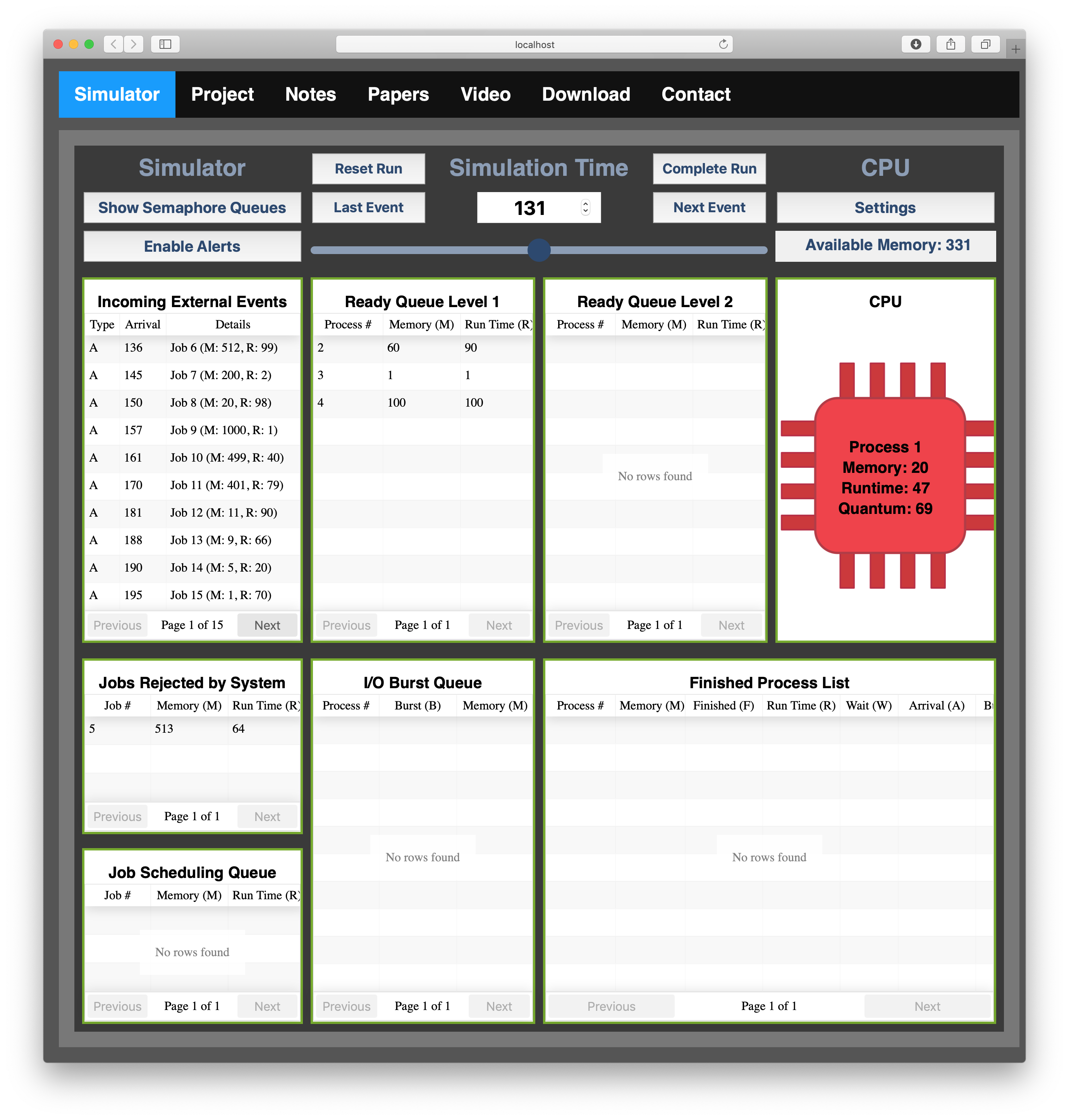}
\end{tabular}}
\caption{Sample screen from the simulation tool (Screen 9 of 17).}
\label{fig:demo9}
\end{figure}

Clicking the `Next Event' button twice more brings the simulation to time 130,
shown in Figure~\ref{fig:demo7}, where two new jobs have been loaded into the
ready queue.  At time 131, there is an incoming external event for the arrival
of job 5.  However, job 5 requires 513 units of memory, which is greater than
the main memory capacity of the system (i.e., 512 units of total memory that
the simulation supports). In this simulation, since there is not enough memory
to accommodate job 5, it can never run and, thus,
is rejected with the alert popup message
shown in Figure~\ref{fig:demo8}.  This presents an opportunity to mention to
students that in a virtual memory management scheme---a forth-coming
topic---this job would be runnable, even though it exceeds the total amount of
system memory. To disable alerts, a user can toggle the button labeled
`Disable Alerts' 
above the incoming external events list (see Figure~\ref{fig:demo7}).
After closing the alert, the
simulation time is 131, shown in Figure~\ref{fig:demo9}, where the rejected job
5 is in the list of jobs rejected by the system. At time 136, job 6, which
requires exactly 512 units of memory, arrives. While this job enters the job
queue (which is in secondary memory), it will not be permitted to enter the
ready queue (which is in main memory) until all processes in main memory have
fully completed (i.e., terminated).  At that point, the full 512 units of
memory are free and available for job 6.

\begin{figure}[h!]
\centering
\resizebox{\textwidth}{!}{
\begin{tabular}{c}
\includegraphics[scale=0.20]{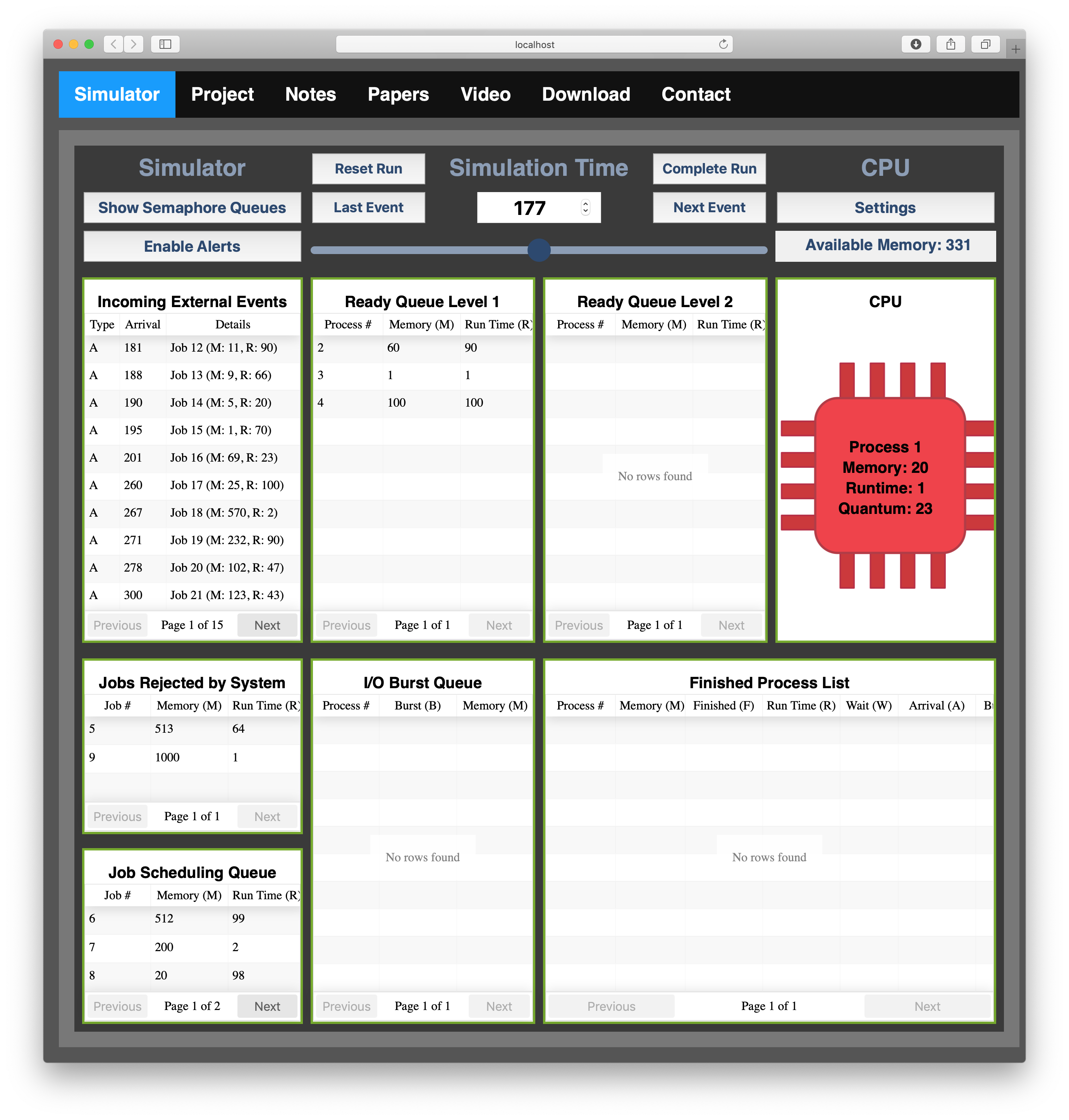}
\end{tabular}}
\caption{Sample screen from the simulation tool (Screen 10 of 17).}
\label{fig:demo10}
\end{figure}

\begin{figure}[h!]
\centering
\resizebox{\textwidth}{!}{
\begin{tabular}{c}
\includegraphics[scale=0.20]{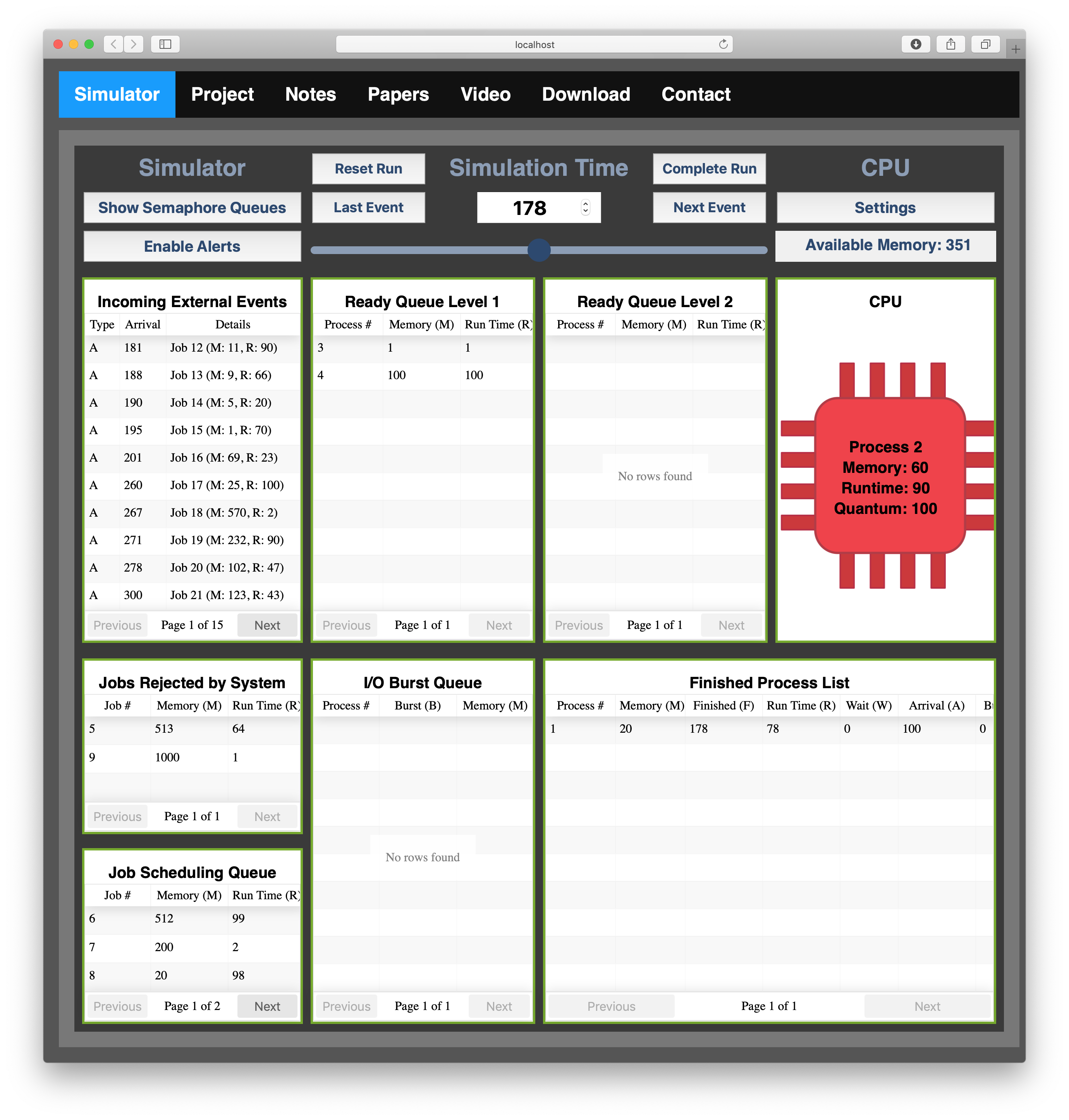}
\end{tabular}}
\caption{Sample screen from the simulation tool (Screen 11 of 17).}
\label{fig:demo11}
\end{figure}

In Figure~\ref{fig:demo10}, the simulation time is now 177 and process 1
requires only 1 unit of time before its completion.  Figure~\ref{fig:demo11}
shows the result of running the simulation for 1 more unit of time after
the button `Next Event' is clicked.
Note that process 1 has moved to the finished process
list, and process 2 has been loaded onto the \textsc{cpu} and requires 90 units
of time to complete.

\begin{figure}[h!]
\centering
\resizebox{\textwidth}{!}{
\begin{tabular}{c}
\includegraphics[scale=0.20]{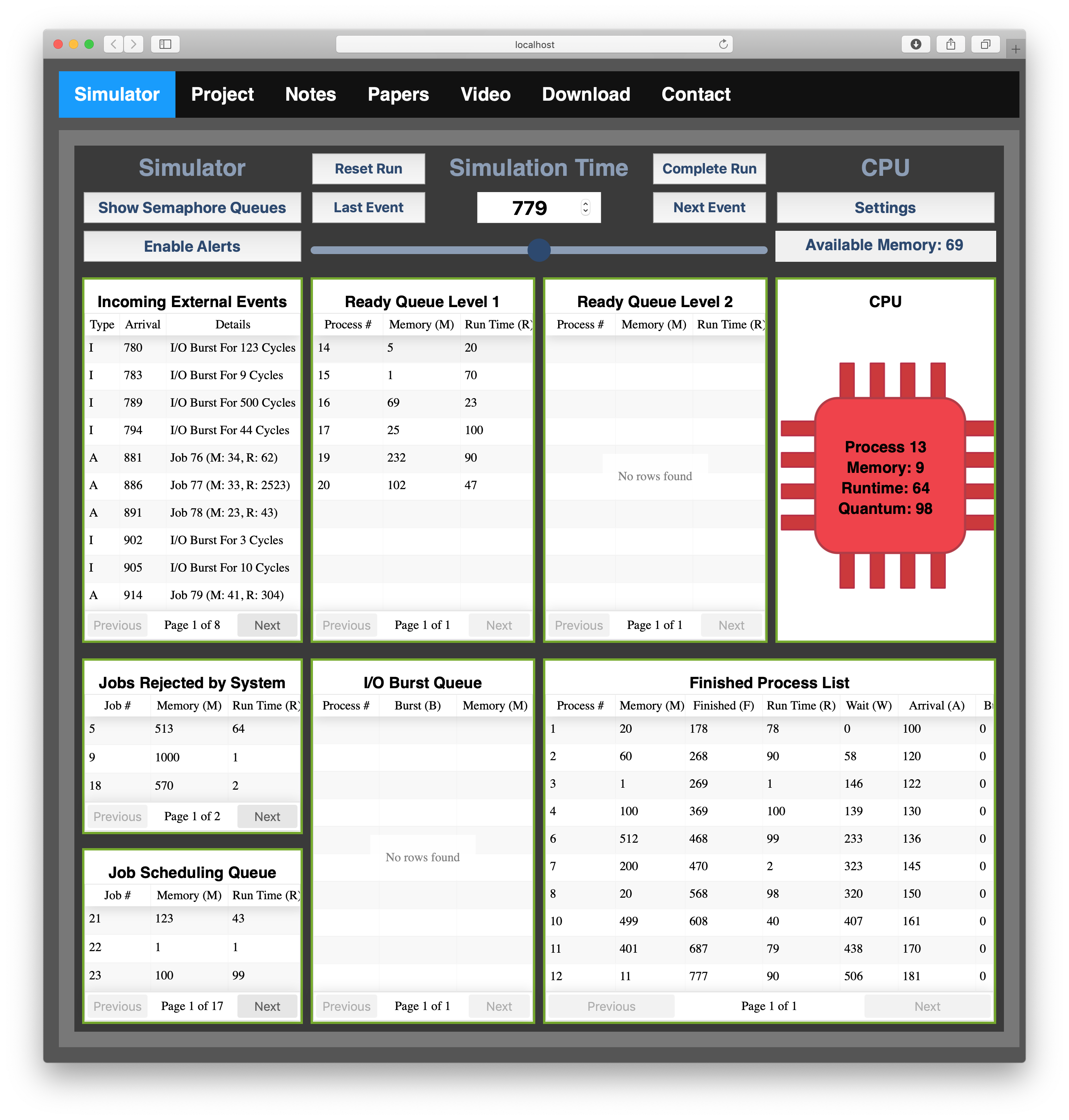}
\end{tabular}}
\caption{Sample screen from the simulation tool (Screen 12 of 17).}
\label{fig:demo12}
\end{figure}

\begin{figure}[h!]
\centering
\resizebox{\textwidth}{!}{
\begin{tabular}{c}
\includegraphics[scale=0.20]{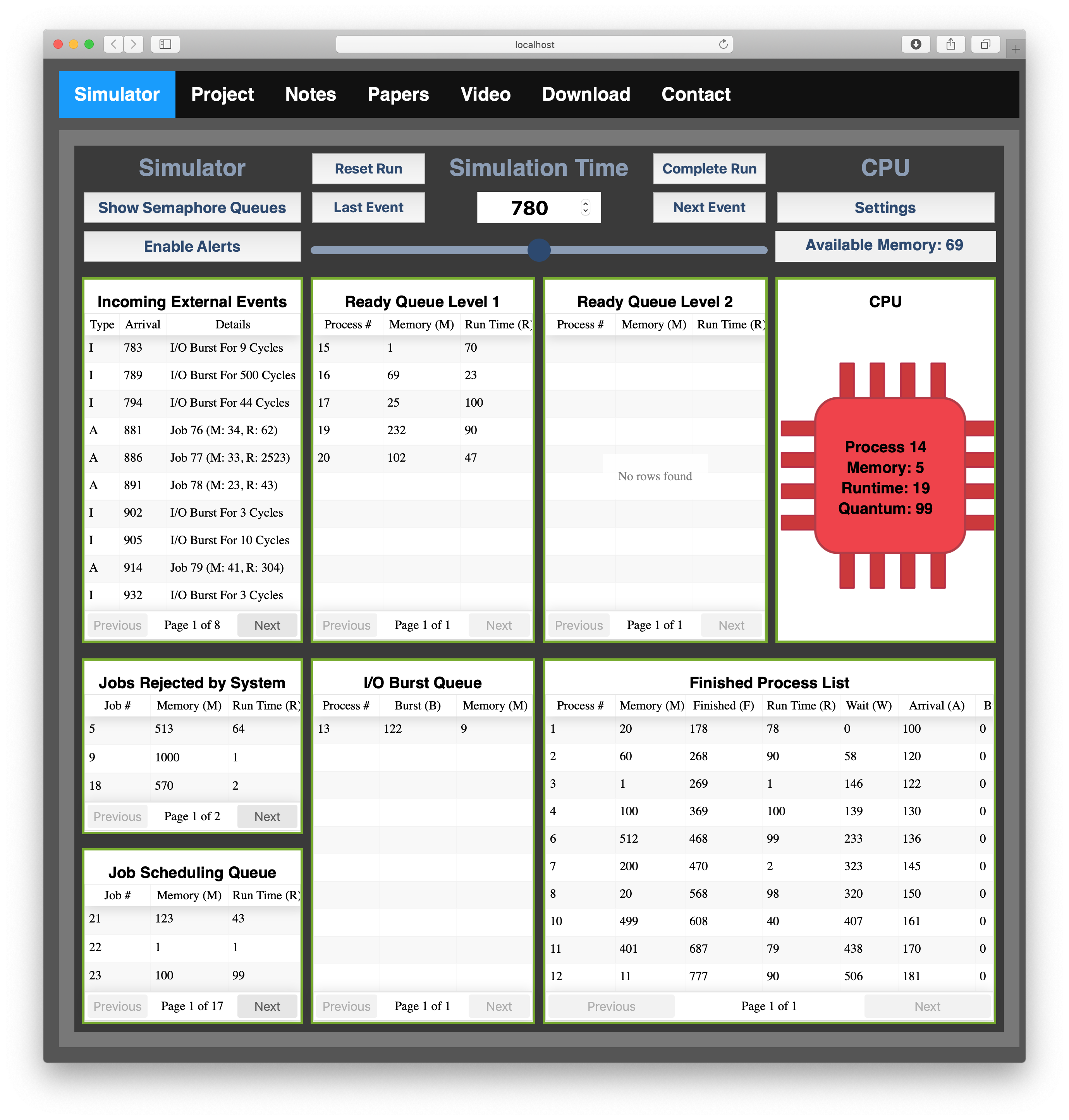}
\end{tabular}}
\caption{Sample screen from the simulation tool (Screen 13 of 17).}
\label{fig:demo13}
\end{figure}

Moving forward to time 779, shown in Figure~\ref{fig:demo12}, we see that
several processes have finished, several are in the first level of the ready
queue, and many jobs are waiting in the job scheduling queue. In the incoming
external events list, we see that the next event, identified with the symbol
`\textsc{i},' occurs at time 780 and is a request for \textsc{i/o} by the
process on the \textsc{cpu}---in this case, process 13. At time 780, shown in
Figure~\ref{fig:demo13}, process 13 leaves the \textsc{cpu} and is moved to the
\textsc{i/o} wait queue for the specified \textsc{i/o} burst. Once the
\textsc{i/o} burst is complete, the process is moved back to the first level of
the ready queue. Many other additional \textsc{i/o} events occur over the next
few time steps. 

\begin{figure}[h!]
\centering
\resizebox{\textwidth}{!}{
\begin{tabular}{c}
\includegraphics[scale=0.20]{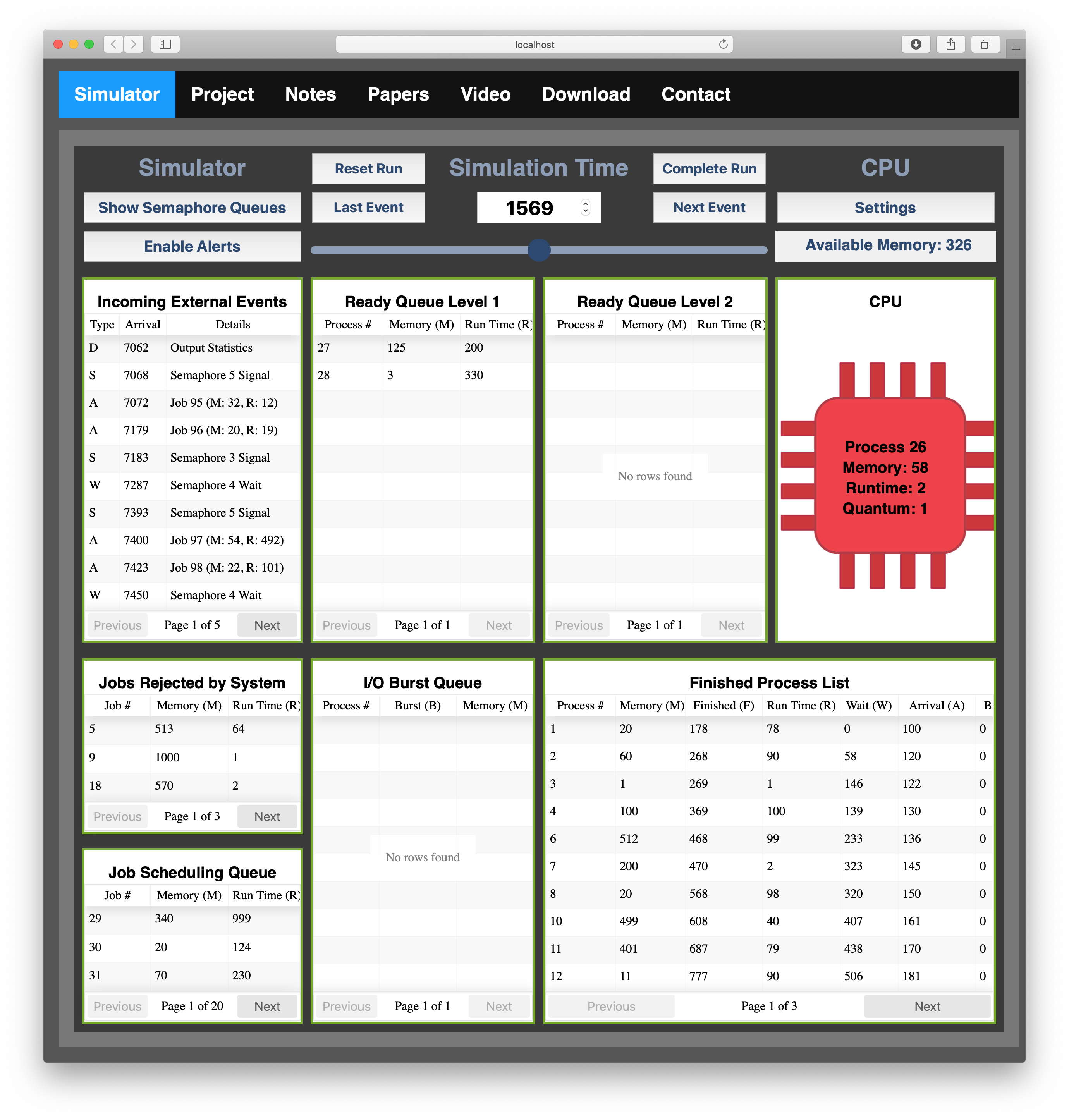}
\end{tabular}}
\caption{Sample screen from the simulation tool (Screen 14 of 17).}
\label{fig:demo14}
\end{figure}

\begin{figure}[h!]
\centering
\resizebox{\textwidth}{!}{
\begin{tabular}{c}
\includegraphics[scale=0.20]{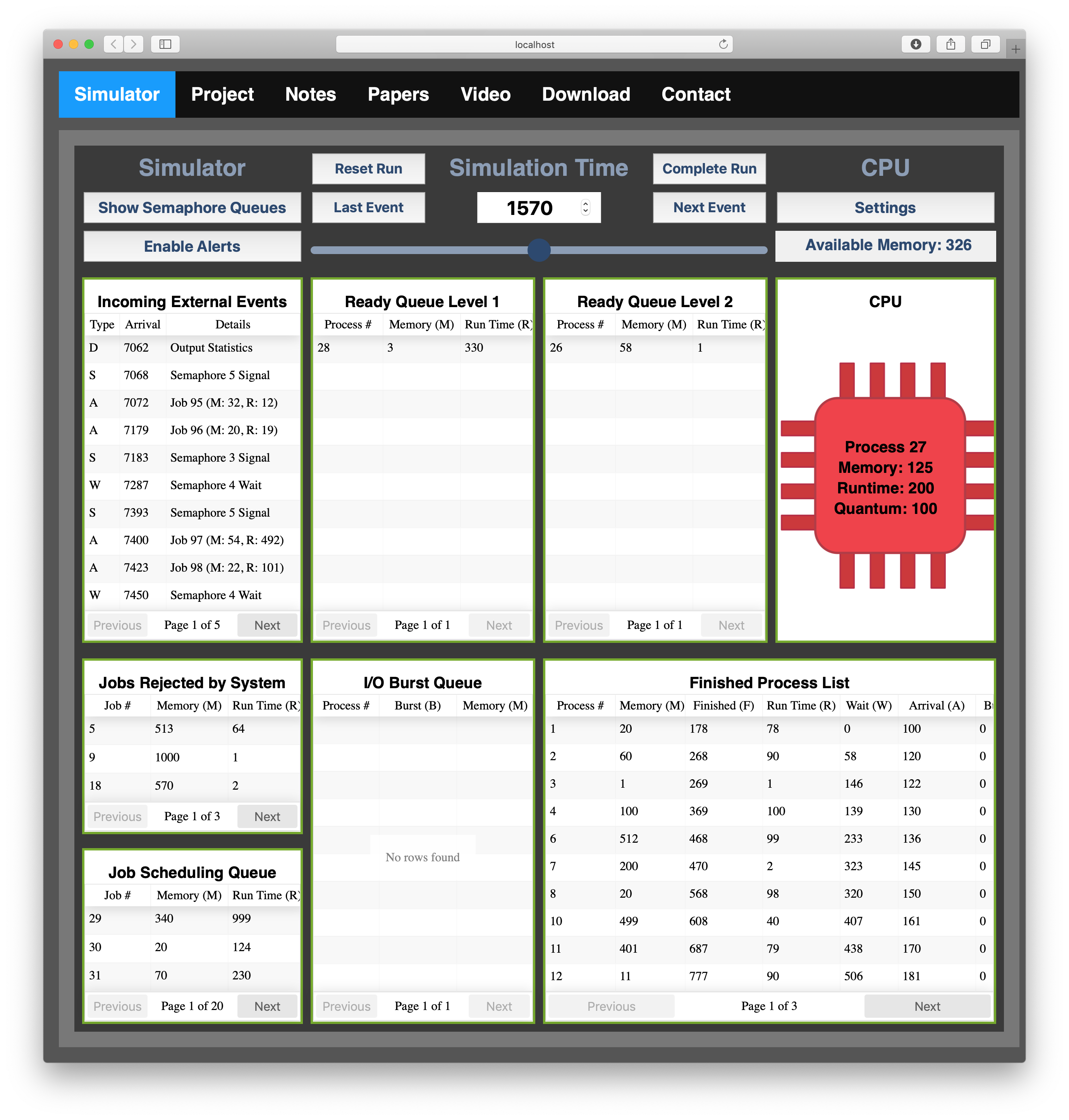}
\end{tabular}}
\caption{Sample screen from the simulation tool (Screen 15 of 17).}
\label{fig:demo15}
\end{figure}

Up to this point, the allotted quantum of 100 units has been sufficient for
each process to finish before a quantum expiration. Thus, the second level
ready queue is yet used.  At time 1,569, shown in Figure~\ref{fig:demo14},
process 26 requires 2 units of time before completion, but its remaining
quantum is only 1 unit of time. In Figure~\ref{fig:demo15}, process 26 is moved
to the second level of the ready queue at time 1,570, and will not get back on
the \textsc{cpu} again until the first level of the queue is empty since
processes on the first level of the ready queue have priority over processes on
the second level of the queue.  While process 26 requires only 1 more unit of
time to complete, it must now wait (potentially indefinitely leading to
starvation due to the priority policy between the levels of the ready queue)
for more time on the \textsc{cpu}. Students often raise questions about the
efficiency of such a scheduling scheme.  We use this opportunity to discuss the
tradeoffs of scheduling algorithms and address potential solutions to
starvation such as aging.

\begin{figure}[h!]
\centering
\resizebox{\textwidth}{!}{
\begin{tabular}{c}
\includegraphics[scale=0.20]{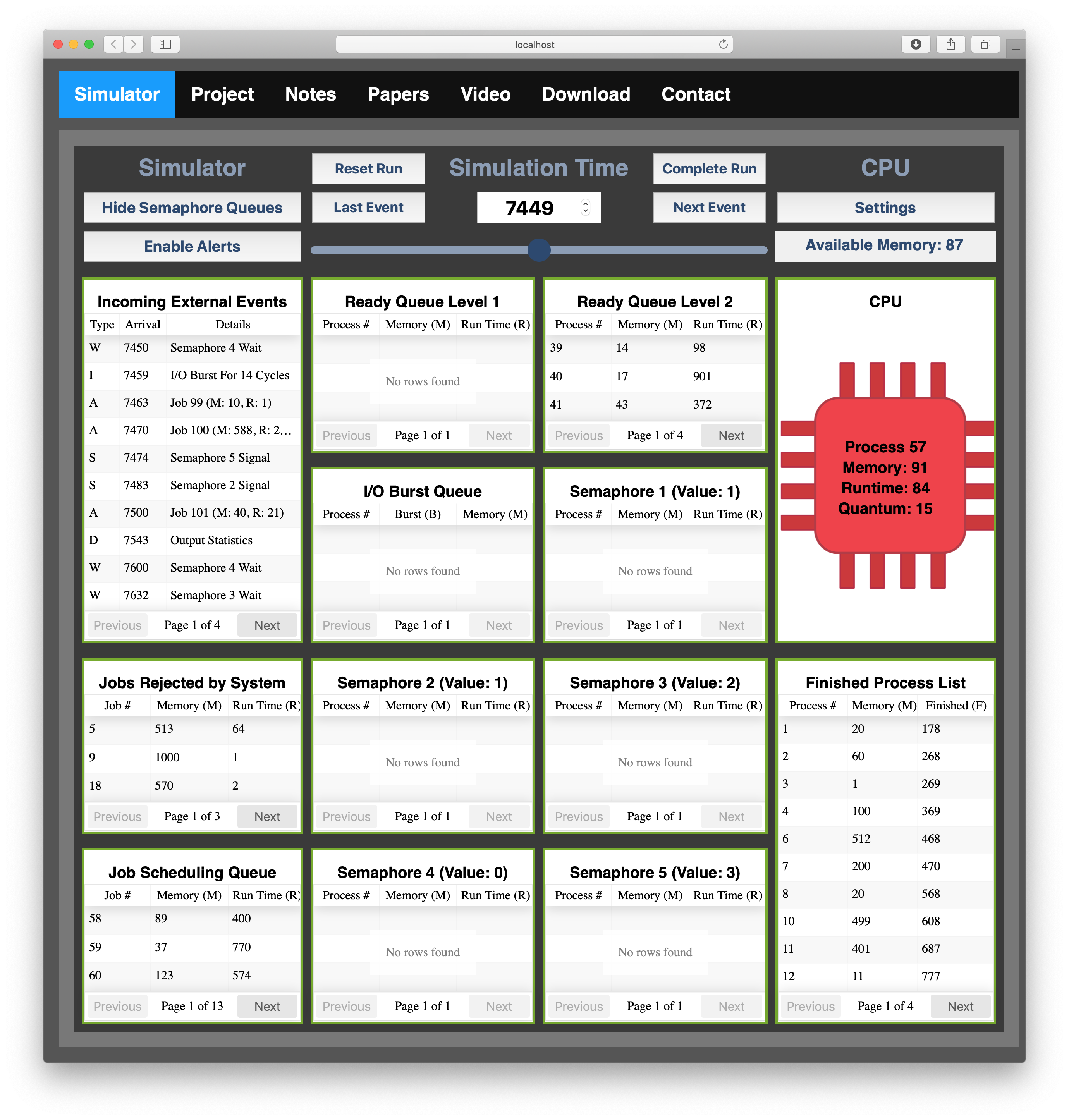}
\end{tabular}}
\caption{Sample screen from the simulation tool (Screen 16 of 17).}
\label{fig:demo16}
\end{figure}

\begin{figure}[h!]
\centering
\resizebox{\textwidth}{!}{
\begin{tabular}{c}
\includegraphics[scale=0.55]{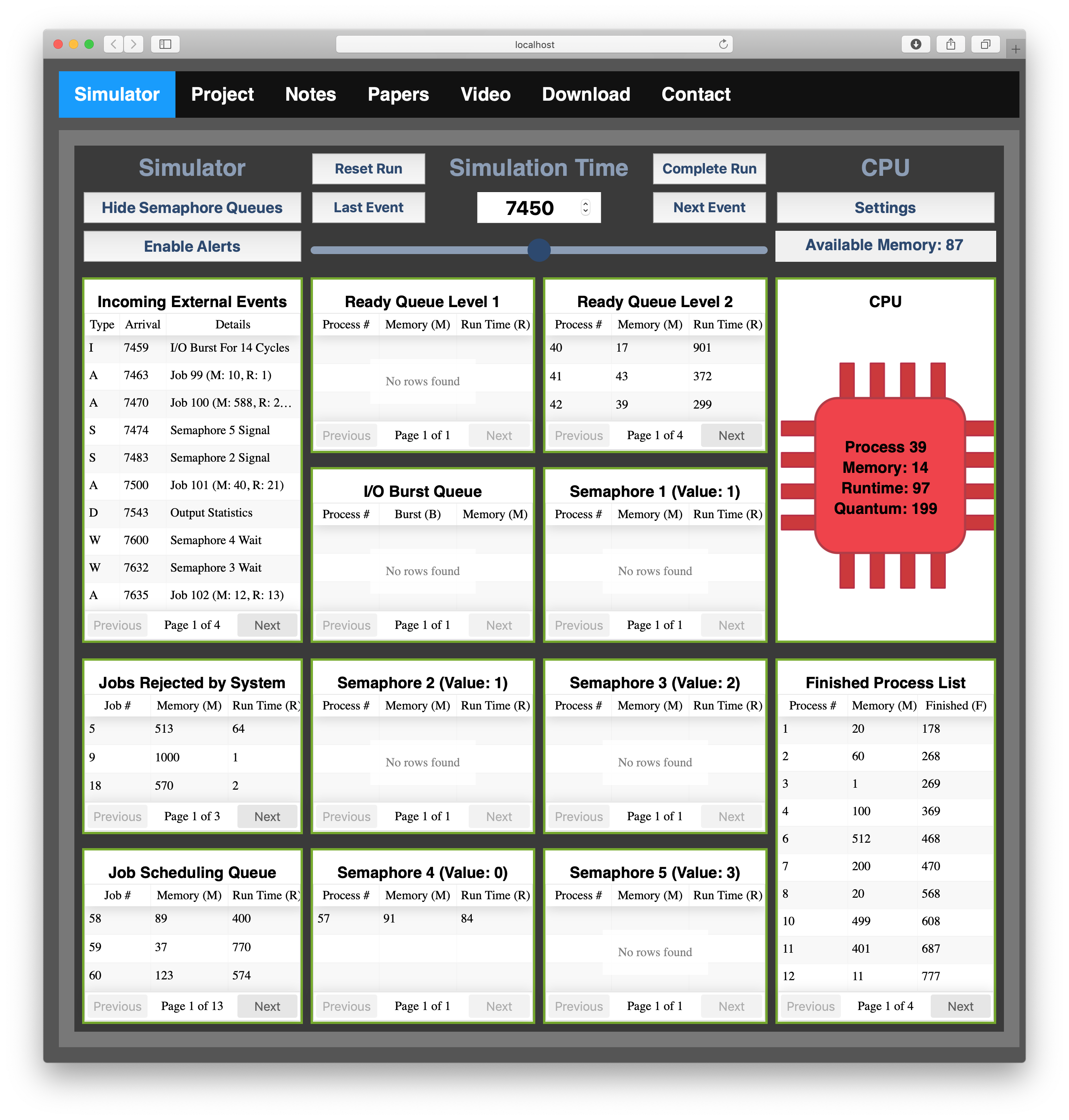}
\end{tabular}}
\caption{Sample screen from the simulation tool (Screen 17 of 17).}
\label{fig:demo17}
\end{figure}

We now demonstrate the use of the system semaphores.  We can toggle the button
labeled `Show Semaphore Queues' to restore the semaphore queues to the display.
The first semaphore event is a signal which is identified by the `\textsc{s}'
symbol in the incoming external events list.  This signal occurs at time 7,068
and signals semaphore 5 (see Figure~\ref{fig:demo15}).  The availability of a
semaphore is tracked next to the semaphore labels in the semaphore wait queues
(see Figure~\ref{fig:demo16}).  When a semaphore wait event, identified with a
`\textsc{w}' symbol, occurs, it causes the process on the \textsc{cpu} to
acquire or wait on the identified semaphore.  If that semaphore is available,
its value is decremented, the process acquires the semaphore, and, thus,
remains on the \textsc{cpu}.  If is that semaphore is unavailable (i.e., has a
value of 0), the process on the \textsc{cpu} must block and, thus, is moved to
the particular semaphore wait queue until the semaphore is available (through a
subsequent signal). At time 7,449, shown in Figure~\ref{fig:demo16}, the
simulation is 1 unit of time away from an incoming external event requiring
process 57 on the \textsc{cpu} to acquire semaphore 4.  Since the value of
semaphore 4 is 0, process 57 blocks, leaves the \textsc{cpu}, and must wait in
the wait queue for semaphore 4 at time 7,450 (see Figure~\ref{fig:demo17}).
For additional sample simulation scenarios, see a YouTube video of a text-based
demonstration of the underlying simulation engine available at
\url{https://youtu.be/eRU8h-5aMOs}.

\section{A Related Tool}

A similar, albeit non-graphical, \textsc{cpu} scheduling simulator is available
at \url{http://classque.cs.utsa.edu/classes/cs3733s2015/notes/ps/index.html} as
a Java applet: \texttt{appletviewer}
\url{http://classque.cs.utsa.edu/classes/cs3733/scheduling2/index.html}. This
tool allows the user to explore a variety of \textsc{cpu} scheduling algorithms
(e.g., \textsc{fcfs}, \textsc{sjf}, \textsc{psjf}, and \textsc{rr}).  The user
chooses an algorithm, sets a quantum, and inputs an arrival time, \textsc{cpu}
burst time, and \textsc{i/o} burst time for each process in a set of
user-defined processes.  The tool then produces a text-based Gantt chart.  When
a change is made to any of these inputs, the Gantt chart is updated.  There are
some key differences between this tool and our tool with implications
on student learning.  While our tool, like this tool, does permit the user
to dynamically tune the quantum, unlike this tool, our tool does not permit the
user to select the scheduling algorithm---the round-robin (\textsc{rr})
scheduling algorithm
is fixed in our tool.  However, and more importantly, unlike this tool, our
tool i) simulates and displays the variety of queues involved in \textsc{cpu}
scheduling and \textsc{i/o} and semaphore processing, ii) supports the user in
stepping through the simulation at user-defined increments of time, and iii)
allows the user to dynamically tune more simulation parameters than just
quantum.  In short, unlike our tool, this related tool does not capture the
transitions from one unit of time to the next.  (It also has an upper bound on
how many processes can be simulated before the textual Gantt chart becomes
unreadable.  Also, the Java web applet is not secure and is blocked by most
modern web browsers by default.)

Our tool allows users to interactively step through the simulation by any
increment of time and observe both the state and location of all processes.
Rather than displaying only the state of processes, our tool also illustrates
the queues in which each process resides throughout its lifecycle.  For example,
in our tool, users can monitor a process as it is loaded into memory, inserting
into the ready queue, granted access to the \textsc{cpu}, preempted to a
semaphore or \textsc{i/o} wait queue, time sliced and moved to the second-level
ready queue, and eventually completed and flushed out of the system onto the
list of finished processes.  Moreover, our tool supports the dynamic tuning of
many of the simulation parameters.  For instance, users can inject or edit
incoming events at any time (e.g., altering the semaphore signals at any
increment of time along the simulation timeline). The ability to step backwards
also allows users a convenient way to explore multiple scenarios from a given
point in time.  This level of interaction supported by our tool provides ample
scope for students to explore \textsc{cpu} scheduling in a variety of
user-created scenarios.  Lastly, unlike this tool, our tool also involves
memory usage and multi-level feedback queues.

\section{Student Feedback and Discussion}

\subsection{Student Comments}

Students across a wide range of offerings of the \textsc{os} course have found
the project to develop the underlying simulation engine helpful for discerning
and acquiring an appreciation of the difficulty in the copious event processing
an \textsc{os} must handle.  It also gives them a feel for the operations
management nature of an \textsc{os}.  The following is a sample of anonymous
student feedback from a course evaluation.

\begin{quote} \textit{The project really nailed in the main concepts of
operating systems in general.} \end{quote}

\begin{quote} \textit{The project was also an interactive and engaging
experience that demonstrated and explained concepts we were working on in
class.}  \end{quote}

\begin{quote} \textit{I found that the project really helped me learn how an
operating system scheduler worked.} \end{quote}

\begin{quote} \textit{Also I found the project to be really fun, I actually
enjoyed working on it.} \end{quote}

\begin{quote} \textit{\dots mostly the project that we did halfway through the
semester was very beneficial to my learning looking back at it.} \end{quote}

Another use of our tool is as an aid to students working on conceptual,
pencil-and-paper
process scheduling exercises (such as those in~\cite{osc10}[Chapter
5]), particularly for verifying the correctness of their work.  In addition to
its use for exploring the demonstrated concepts, we have discovered an
unintended use of this graphical tool---students can use it as a tool to debug
their underlying simulation engine.  Observing the operation of their
underlying simulation graphically helps students identify bugs in their
implementation more quickly than wading through pages of textual dumps of their
system queues, e.g., to identify a process that went awry.

Approximately three hundred and twenty students have completed the underlying
simulation engine project since the Fall 2009 semester. Students are permitted
to use any programming language of their choice for implementation.  Students
have primarily used Java (210) and C++ (95)---the languages used in our
introductory sequence.\footnote{We switched from C++ to Java in Fall 2014, but
C++ was phased out progressively over the span of a few semesters in the
three-course sequence.} Students have also used Python (12), C$^{\sharp}$ (2),
and Perl (1).  Two students re-implemented their projects---one in Scheme and
one in Elixir.  (The Elixir program was multi-threaded---see below.)

There are multiple extensions to the underlying simulation engine that can be
assigned to students as follow-on projects.  For instance, students can
implement a memory management scheme (e.g., paging) to the organization of the
ready queues and/or simulate a multi-core processor.  Also, note that the
underlying simulation engine is a single-threaded program, albeit one that
simulates multiple concurrent processes.  Once students implement the
single-threaded implementation, a possible segue into the topics of concurrency
and synchronization is to re-design/implement the simulator in a multi-threaded
fashion using an event-based concurrency model such as the Actor model of
concurrency in, e.g., Elixir.  One student did this for extra credit.

\subsection{Download and Install Locally}

The entire simulation tool can be downloaded and run locally. The tool requires
\texttt{Node.js} version 10 or 11 and npm version 6 to be installed. A
production build of the tool can be downloaded from
\url{https://cpudemo.azurewebsites.net/#/download}.  Complete up-to-date
instructions are also available with the download.

\subsection{Future Work}

We plan multiple enhancements to the graphical overly intended to make the tool
more flexible to facilitate its use by other educators. For instance, we plan
to compute and display a variety of performance metrics at the completion of
the simulation (e.g., average turnaround time and average job scheduling wait
time).  We would also like to support other scheduling algorithms (e.g.,
\textsc{sjf}, \textsc{psjf}).  We also plan to redesign the event queue so that
it is editable enabling an instructor to dynamically alter the simulation
scenario.  Recall, that multiple canned sequences of incoming external events
are currently available that demonstrate many of the aspects of the simulation.
Ultimately, we plan to decouple our graphical visualization from the underlying
text-based simulator.  This will allow an instructor to overlay our
visualization on top of any operating system for which the instructor's
students are simulating.  This will afford the instructor the freedom to
customize the design requirements and parameters of the particular operating
system the students are simulating while still being able to make use of the
visualization for purposes of any of its pedagogical purposes.  Moreover, this
decoupling will enable students make use of the graphical overlay as a tool to
debug their underlying simulation engine.  Observing the operation of their
underlying simulation graphically will help students identify bugs in their
implementation more quickly than wading through pages of textual dumps of their
systems queues (e.g., to identify a process that went awry).

This software simulation/demonstration is part of a NSF-funded project
whose goal is to foster innovation, in both content and delivery, in teaching
\textsc{os} through the development of a contemporary model for an \textsc{os}
course that aims to resolve issues of misalignment between existing
\textsc{os} courses and employee professional skills and knowledge
requirements.

\newpage

\newpage

\newpage

\newpage

\newpage

\newpage

\newpage

\newpage

\newpage

\newpage

\newpage

\newpage

\newpage

\newpage

\newpage

\newpage

\section*{Acknowledgments}

This material is based upon work supported by the National Science Foundation
under Grant Numbers 1712406 and  1712404.  Any opinions, findings, and
conclusions or recommendations expressed in this material are those of the
author(s) and do not necessarily reflect the views of the National Science
Foundation.  The source of the underlying simulation engine is a course project
designed by John A. Lewis in which students design and implement a program that
simulates some of the job and \textsc{cpu} scheduling, and semaphore processing
of a time-shared operating system.

\end{document}